\documentclass[epj]{svjour}
 
\usepackage{amsmath,amssymb,graphicx,enumitem,multirow,color}

\begin{document}

\title{Combination of analysis techniques for efficient track
reconstruction in high multiplicity events} 

\author{Ferenc Sikl\'er}
 
\institute{Wigner Research Centre for Physics, Budapest, Hungary}

\date{Received: date / Revised version: date}

\titlerunning{Combination of analysis techniques for track reconstruction in
high multiplicity events}
 
\abstract{
A novel combination of established data analysis techniques for reconstructing
all charged-particle tracks in high energy collisions is proposed. It uses all
information available in a collision event while keeping competing choices open
as long as possible.
Suitable track candidates are selected by transforming measured hits to a
binned, three- or four-dimensional, track parameter space. It is accomplished
by the use of templates taking advantage of the translational and rotational
symmetries of the detectors. Track candidates and their corresponding hits, the
nodes, form a usually highly connected network, a bipartite graph, where we
allow for multiple hit to track assignments, edges. The graph is cut into very
many minigraphs by removing a few of its vulnerable components, edged and nodes. Finally the
hits are distributed among the track candidates by exploring a deterministic
decision tree. A depth-limited search is performed maximising the number of
hits on tracks, and minimising the sum of track-fit $\chi^2$.
Simplified models of LHC silicon trackers, as well as the relevant physics
processes, are employed to study the performance (efficiency, purity, timing)
of the proposed method in the case of single or many simultaneous proton-proton
collisions (high pileup), and for single heavy-ion collisions at the highest
available energies.
 \PACS{
  {29.40.Gx}{Tracking and position-sensitive detectors} \and
  {29.85.-c}{Computer data analysis}
 }
}

\maketitle

\def\pt{p_\text{T}}
\def\GeVc{Ge\hspace{-.08em}V\hspace{-0.16em}/\hspace{-0.08em}$c$}
\def\MeV{\text{Me\hspace{-.08em}V}}
\def\mum{\mathrm{\mu}m}

\def\q_R{k_\text{T}}

\section{Introduction}

Traditional methods of track reconstruction can be scaled to work in high
multiplicity events, namely in many simultaneous collisions (pileup) of
elementary particles~\cite{Aad:2015ina,Rovere:2015rep} and in high multiplicity
single heavy-ion collisions. Nevertheless the performances are not optimal,
efficiency and purity are reduced, especially at low momentum.
That is why present data taking conditions and further luminosity and energy
upgrades of high energy particle colliders, as well as those of detector
systems, call for new ideas.

Image transformation methods and neural networks~\cite{Fruhwirth:1993wv} are
often used in gaseous detectors (time projection
chambers~\cite{Aamodt:2008zz,Cheshkov:2006ym} and transition radiation
trackers~\cite{Aad:2008zzm,Mindur:2017nqn}). In the case of silicon trackers
the combinatorial track finding methods employed for trajectory building mostly
use local information~\cite{Strandlie:2004ig,Chatrchyan:2008aa}. They start
with a trajectory seed and build the trajectory by extending the seed through
the detector layers, picking up compatible hits. In the case of very many
compatible hits the number of concurrently built trajectory candidates must be
limited. Only some of the best candidates are kept which biases the final
result. In this sense, decisions are made too early. Moreover, trajectories are
mostly treated separately, there is no interaction between their assigned hits.

In this study a combination of established data analysis techniques for the
offline reconstruction of all charged-particle trajectories is proposed. It
uses all information available in an event while keeping competing choices open
as long as possible.
Details of silicon detectors, relevant physical effects and tracking with
Kalman filter are introduced in Sec.~\ref{sec:two}. Pattern recognition along
with the preparation of templates, image transformation, and trajectory
building are discussed in Sec.~\ref{sec:image}. The optimal distribution of
hits among tracks with help of graph-theoretic methods are shown in
Sec.~\ref{sec:graph}.
Results of simulations based on simplified but realistic models of silicon
trackers at particle colliders are detailed in Sec.~\ref{sec:results}, where
the performance (efficiency, purity, timing, parallelisation) of the proposed
methods are displayed.

\begin{table*}[!t]

 \caption{The main characteristics of the inner barrel silicon detectors of
the studied experimental setups. The value of the longitudinal magnetic field
$B_z$ is shown along with layer type, radii of barrel cylinders, tilt angle (in
case of double-sided strips), spatial resolution of hits in $r\phi$ and $z$
directions, and material thickness $x/X_0$ in radiation length units. In the
case of strips the strip length $l_z$ is given instead, in parentheses.}

 \label{tab:setups}

 \begin{center}
 \begin{tabular}{cccccccc}
  \hline
   & $B_z$ [T] & Layer type & Radii [cm] & Tilt [mrad]
   & $\sigma_{r\phi}$ & $\sigma_z$ ($l_z$) & $x/X_0$ [$\%$] \\
  \hline
  \multirow{4}{*}{\rotatebox{90}{Exp A}} & \multirow{4}{*}{2.0}
    & pixels & 5.0, 8.8, 12.2 & --    & 10~$\mum$ & 115~$\mum$ & 4 \\
  & & strips & 29.88, 29,92    & $\pm20$ & 17~$\mum$ & (6.4~cm) & 2 \\
  & & strips & 37.08, 37,12    & $\pm20$ & 17~$\mum$ & (6.4~cm) & 2 \\
  & & strips & 44.28, 44,32    & $\pm20$ & 17~$\mum$ & (6.4~cm) & 2 \\
  \hline
  \multirow{4}{*}{\rotatebox{90}{Exp B}} & \multirow{4}{*}{0.4}
    & pixels &  3.9,   7.6  & --      & 12~$\mum$ & 100~$\mum$ & 1 \\
  & & drifts & 14.9,  23.8  & --      & 35~$\mum$ &  25~$\mum$ & 1 \\
  & & strips & 38.48, 38.52 & $+7.5, -27.5$ & 20~$\mum$ & (4~cm) & 0.5 \\
  & & strips & 43.58, 43.62 & $+7.5, -27.5$ & 20~$\mum$ & (4~cm) & 0.5 \\
  \hline
  \multirow{5}{*}{\rotatebox{90}{Exp C}} & \multirow{5}{*}{3.8}
    & pixels &  4.4, 7.3, 10.2 & --      & 15~$\mum$ & 15~$\mum$ & 3 \\
  & & strips & 25.48, 25,52    & $\pm50$ & 23~$\mum$ & (10~cm) & 2 \\
  & & strips & 33.88, 33.92    & $\pm50$ & 23~$\mum$ & (10~cm) & 2 \\
  & & strips & 41.8            & 0       & 35~$\mum$ & (10~cm) & 2 \\
  & & strips & 49.8            & 0       & 35~$\mum$ & (10~cm) & 2 \\
  \hline
 \end{tabular}
 \end{center}

\end{table*}

\section{Silicon detectors at particle colliders}
\label{sec:two}

At currently operating particle colliders the interaction region is very narrow
(of the order of $50\,\mum$) in transverse direction, while in $z$ (longitudinal
or beam) direction it is long, with a characteristic size of about
10~cm~\cite{MLamont}.
For single heavy-ion collisions the $z_0$ position of the primary interaction
(vertex) is estimated with good precision using the copiously produced high
transverse-momentum particles, thanks to the small pointing uncertainty of
their tracks, reconstructed with traditional methods. In the case of single or
multiple pp collisions no such information on the locations of the interaction
vertices exists.

The trajectory of a primary particle is primarily determined by its initial
position $(0,0,z_0)$ and parameters $(q, \eta, \pt, \phi_0)$ of its initial
momentum at creation. Here $q$ is the charge, $\eta = -\ln\tan(\theta_0/2)$ is
the pseudorapidity, $\pt$ is the transverse momentum, $\phi_0$ and $\theta_0$
are the azimuthal and polar angles of the initial momentum vector in spherical
coordinates.
In a large volume solenoid the magnetic field near the center of the detector
is rather homogeneous and points in the $z$ direction. Hence in small volumes
the trajectories of charged particles can be approximated by piecewise helices.
For practical purposes a primary particle is parametrised by $(\q_R, \sinh\eta,
\phi_0, z_0)$ in the following, where $\q_R = q/R$ is the signed curvature of
the projection of its trajectory on the transverse (bending) plane, $R$ is its
radius. If the particle is singly charged the curvature is connected to $\pt$
as $\pt = e B_z R$, where $e$ is the electric charge of a proton, $B_z$ is the
value of the longitudinal magnetic field.

The central parts of silicon trackers generally consist of several concentric
cylindrical layers. Those close to the nominal interaction point are equipped
with tiny pixel sensors, while others contain long strip sensors parallel with
the beam direction. Some strip layers are double-sided, they are located very
close to each other two by two, and have a small relative tilt angle.
The main characteristics of the inner barrel silicon detectors of the studied
experimental setups are given in Table~\ref{tab:setups}.

\begin{figure*}[!t]

 \begin{center}
  \resizebox{\textwidth}{!}{
   \input{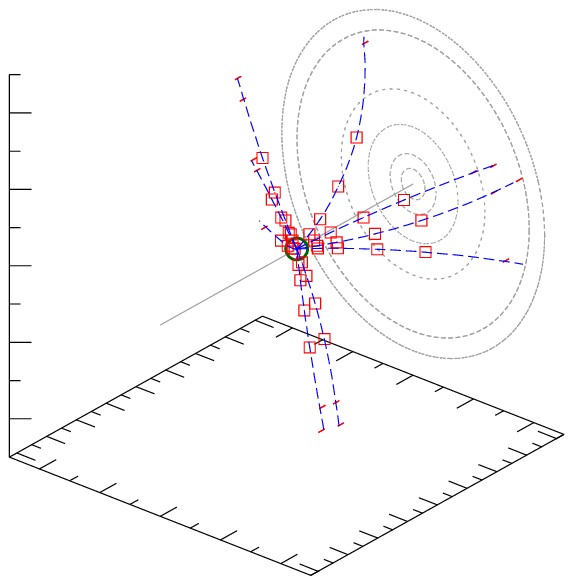}
   \input{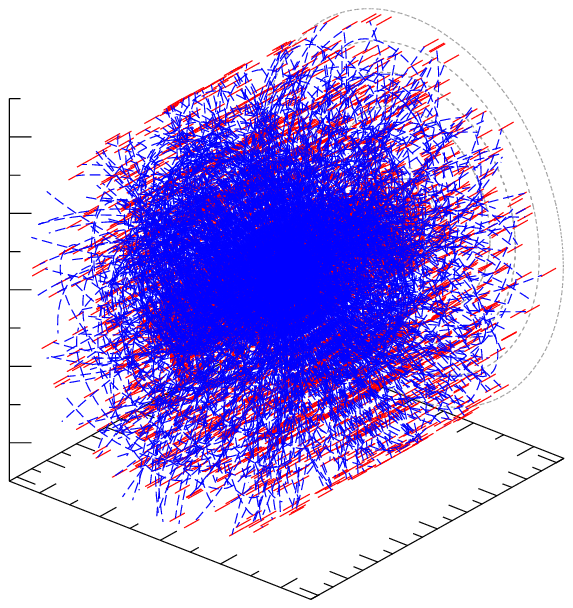}
  }
 \end{center}

 \caption{Left: pixel hits (red open squares) and strip hits (red solid
line sections) of a single inelastic pp collision from the simulation of Exp B.
The location of the primary interaction is plotted with a green circle, the
beamline is indicated by a gray straight line. Charged particle trajectories
(blue dashed curves) are also plotted.
Right: event with 40 simultaneous inelastic pp collisions from the simulation
of Exp C.}

 \label{fig:events_full}

\end{figure*}

The trajectory of the primary particle intersects the concentric cylindrical
layers and leaves hits behind in the silicon (Fig.~\ref{fig:events_full}). In
the simplified case, when the magnetic field is homogeneous and if the detector
material and its physical effects are neglected, the position of those hits
could be precisely determined by simple equations. The physical effects of
detector material changes this overly simple picture.

\subsection{Physical effects}
\label{sec:physics}

When a long lived charged particle propagates through material the most
important effects which alter its momentum vector are multiple scattering and
energy loss.
The distribution of multiple Coulomb scattering is roughly
Gaussian~\cite{Olive:2016xmw}, the standard deviation of the planar scattering
angle is
\begin{equation}
 \theta_0 = \frac{13.6~\MeV}{\beta c p} z \sqrt{x/X_0}
  \bigl[1 + 0.038 \ln(x/X_0) \bigr],
 \label{eq:theta_0}
\end{equation}

\noindent where $p$, $\beta c$, and $z$ are the momentum, velocity, and charge
of the particle in electron charge units, and $x/X_0$ is the thickness of the
scattering material in radiation lengths.

Momentum and energy is lost during traversal of sensitive detector layers and
support structures. To a good approximation the most probable energy loss
$\Delta_p$, and the full width of the energy loss distribution at half maximum
$\Gamma_\Delta$~\cite{Bichsel:1988if} are
\begin{align}
 \label{eq:mp_energyLoss}
 \Delta_p &= \xi \left[\ln\frac{2 mc^2 \beta^2 \gamma^2 \xi}{I^2}
             + 0.2000 - \beta^2 - \delta \right], \\
 \label{eq:gamma_energyLoss}
 \Gamma_\Delta &= 4.018 \xi,
\end{align}

\noindent where
 $\xi = \frac{K}{2} z^2 \frac{Z}{A} \rho \frac{x}{\beta^2}$
is the Landau parameter; $K = 4\pi N_A r_e^2 m_e c^2$; $m$ is the mass of the
particle; $Z$, $A$, $I$, and $\rho$ are the mass number, atomic number,
excitation energy, and the density of the material,
respectively~\cite{Olive:2016xmw}. The density correction $\delta$ is
neglected.

\subsection{Hit clusters}
\label{sec:clusters}

An incoming charged particle loses energy in the sensitive detector
elements by producing electron-hole pairs. The neighboring channels collecting
a charge above a given threshold are grouped to form a cluster, a reconstructed
hit. The size (dimensions) of the cluster depends on the angle of incidence
of the particle: bigger angles result in larger clusters.
The expected cluster dimensions in $r\phi$ and $z$ directions in pitch units
are
\begin{align*}
 m_{r\phi} &= \frac{t_r |\tan\psi|}{t_{r\phi}}, & 
 m_z       &= \frac{t_r |\tan\theta|}{t_z},
\end{align*}

\noindent where $\theta$ and $\psi$ are local angles (Sec.~\ref{sec:kalman}),
$t_r$ is the thickness of the layer in radial direction, while $t_{r\phi}$ and
$t_z$ are the dimensions of sensitive elements, pitches, in azimuthal ($r\phi$)
and longitudinal ($z$) directions, respectively.
For simplicity, the values $t_r = 300\,\mum$, $t_{r\phi} = 100\,\mum$,
and $t_z = 200\,\mum$ are chosen in the simulation (Sec.~\ref{sec:results}) for
each experimental setup.
Due to the large fluctuations
in energy loss, the measured $w_{r\phi}$ and $w_z$ dimensions of the clusters
differ from the expected ones. In order to model these effects in the
simulation, the cluster dimensions are varied by one unit in both directions
for pixels, and up and down by two units in $r\phi$ direction for strips.

\begin{figure}[!t]

 \begin{center}
  \resizebox{0.9\linewidth}{!}{
   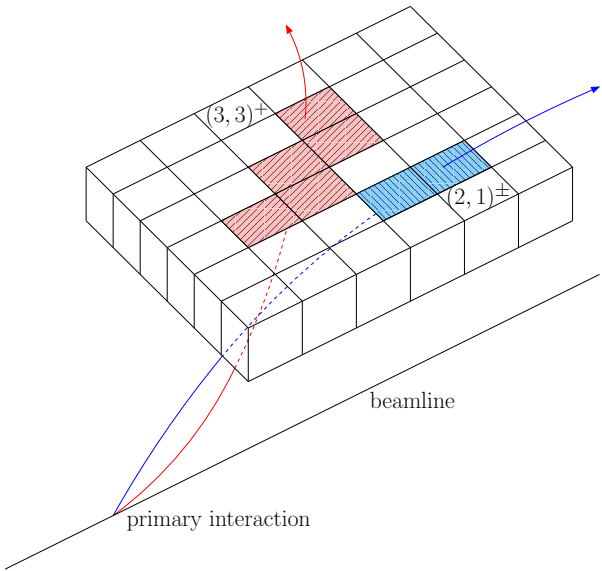
  }
 \end{center}

 \caption{Illustration of the connection between the hit cluster shape and the
local momentum vector, as well as the electric charge of the particle. The
low-momentum positively charged particle (red) leaves a cluster with dimensions
$(3,3)$, in pitch units, in the pixel detector. The fast charged particle (blue)
induces a smaller $(2,1)$ cluster; in this case the electric charge cannot
be unambiguously determined.}

 \label{fig:cluster}

\end{figure}

If the size of a pixel cluster is at least two units in both directions, the
layout of its pixels with charge deposit and its location relative to the
nominal interaction point usually indicate the sign of the electric charge of
the particle (Fig.~\ref{fig:cluster}). This way a pixel cluster is
characterised by the measured widths $w_{r\phi}$ and $w_z$ (dimensions of its
rectangular envelope), and the charge $q$, which can be $1$ or $-1$, or left
unknown. A strip cluster has only one such quantity, its $w_{r\phi}$ width.

\subsection{Particle tracking}
\label{sec:kalman}

The Kalman filter is widely used in particle physics experiments for charged
track and vertex finding and fitting, and provides a coherent framework for
handling known physical effects and measurement
uncertainties~\cite{Fruhwirth:1987fm}. It is equivalent to a global linear
least-squares fit which takes into account all correlations coming from process
noise. It is the optimum solution since it minimises the mean square estimation
error.

The state vector $\vec{x} =
(\kappa, \theta, \psi, r\phi, z)$ is five dimensional:
\begin{align*}
 \kappa &= q/p             & \text{(signed inverse momentum)}, \\
 \theta &= \theta(\vec{p}) & \text{(local polar angle)}, \\
 \psi   &= \phi({\vec p})  & \text{(local azimuthal angle)}, \\
 r\phi  &= r\phi({\vec r}) & \text{(global azimuthal position)}, \\
 z      &= r_L             & \text{(global longitudinal position)}.
 \label{eq:kalman}
\end{align*}

\noindent The propagation function $\vec{f}(\vec{x})$ from layer to layer is
calculated analytically using a helix model. Multiple scattering and energy
loss in tracker layers is implemented with their Gaussian approximations shown
in Eqs.~\eqref{eq:theta_0}--\eqref{eq:gamma_energyLoss}. The propagation matrix
$F = \partial \vec{f} / \partial \vec{x}$ is obtained by numerical derivation.
The measurement vector for pixel hits $\vec{m} = (r\phi, z)$ is two
dimensional, for strip hits $\vec{m} = (r\phi)$ it is one dimensional. The
measurement operator for pixels is
\begin{equation*}
  H = \begin{pmatrix}
       0 & 0 & 0 & 1 & 0 \\ 0 & 0 & 0 & 0 & 1
      \end{pmatrix}.
\end{equation*}

\noindent while for strips it is
\begin{equation*}
  H = \begin{pmatrix}
       0 & 0 & 0 & 1 & -\tan\alpha
      \end{pmatrix}.
\end{equation*}

The covariance of the process noise $Q$ is
\begin{gather*}
 Q = (F_\kappa \otimes F_\kappa^T) \sigma_\kappa^2 +
     (F_\theta \otimes F_\theta^T) \sigma_\theta^2 +
     (F_\psi   \otimes F_\psi^T)   \sigma_\psi^2
\end{gather*}

\noindent where $\sigma_\kappa = \kappa \sigma_\Delta/\beta$, $\sigma_\theta =
\sigma_\psi = \theta_0$ and $F_a = \partial \vec{f}/\partial x_a$ is a vector.
Multiple scattering contributes both to the variation of $\theta$
and $\psi$, while energy loss affects only $\kappa$.

The covariance of measurement noise $V$ for pixels is
\begin{equation*}
 V = \begin{pmatrix}
      \sigma_{r\phi}^2 & 0 \\
      0 & \sigma_{z}^2
     \end{pmatrix}.
\end{equation*}

\begin{figure*}[!t]

 \begin{center}
  \resizebox{\textwidth}{!}{
   \input{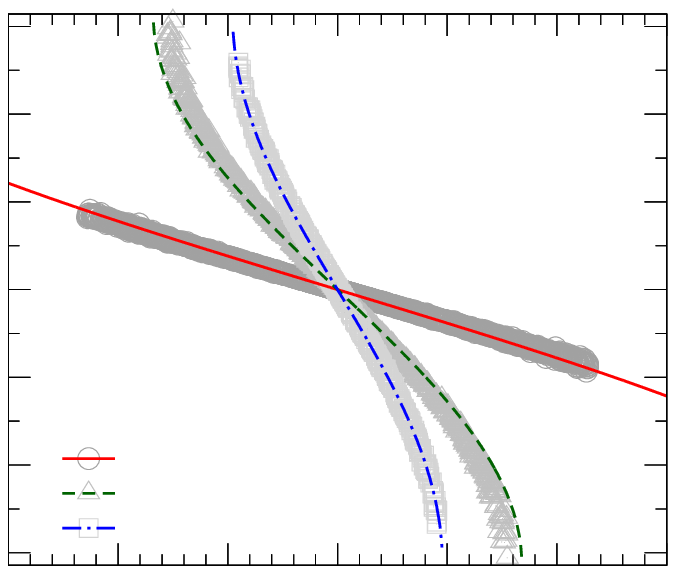}
   \input{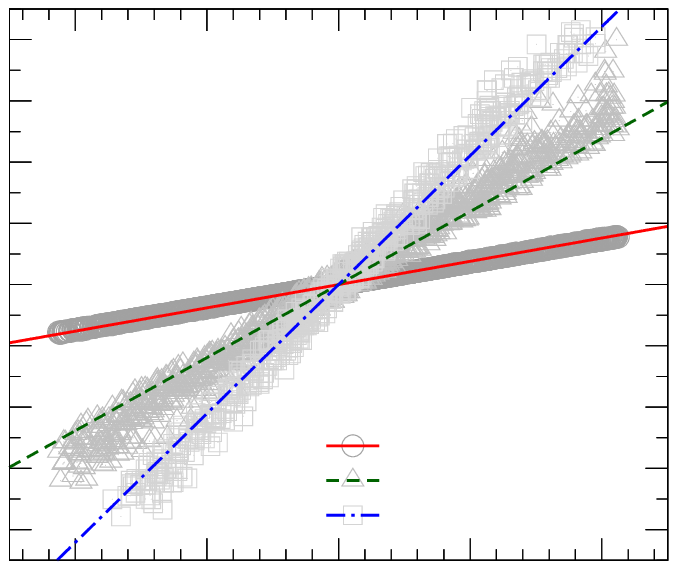}
  }
 \end{center}

 \caption{Left: distributions of $\phi-\phi_0$ differences of the hit and track
azimuth angles as a function of $\q_R$ for some selected detector layers.
Right: distributions of $z-z_0$ differences of the hit and track longitudinal
coordinates as a function of $\sinh\eta$ for some selected detector layers.
Values from simulation of Exp~C are plotted with various symbols, while the
oversimplified expectations $\phi-\phi_0 \approx - \arcsin(r/2 \cdot \q_R)$ and
$z-z_0 \approx 2 \sinh\eta \cdot \arcsin(r/2 \cdot \q_R)/\q_R$ at $\q_R =
0.01~\mathrm{cm^{-1}}$ are shown with the curves.}

 \label{fig:calib_coll}

\end{figure*}

\noindent
In the case of strips with $\alpha$ tilt angle, the inverse of the covariance matrix is
\begin{equation*}
 V^{-1} =
 R(\alpha)
 \begin{pmatrix}
  1/\sigma_{r\phi}^2 & 0 \\
  0 & 0
 \end{pmatrix}
 R^{T}(\alpha) \approx
 \frac{1}{\sigma_{r\phi}^2}
 \begin{pmatrix}
  1 - \alpha^2 & - \alpha \\
  - \alpha & \alpha^2
 \end{pmatrix}
\end{equation*}

\noindent
where $R(\alpha)$ is a rotation matrix.

\begin{table}[!b]

 \caption{Ranges and the optimised number of bins (working point) corresponding
to track parameters. The value of $p_\text{T,min}$ is 0.1~\GeVc.}

 \label{tab:bins}

 \begin{center}
 \begin{tabular}{lcc}
  \hline
  Variable & Range & Bins \\
  \hline
  $\q_R$      & $[- e B_z/p_\text{T,min}, e B_z/p_\text{T,min}]$ &  50 \\
  $\sinh\eta$ & $[-\sinh 1.5, \sinh 1.5]  $                      & 100 \\
  $\phi_0$    & $[-\pi, \pi]$                                    & 200 \\
  $z_0$       & $[-3\sigma_z, 3\sigma_z]$                        &  50 \\
  \hline
 \end{tabular}
 \end{center}

\end{table}

Simulated particles are tracked while they are in the volume of the tracker
detector, that is, trajectories looping in the magnetic field are properly
followed. In the case of pattern recognition and track reconstruction only
inside-out propagation is considered.

\section{Pattern recognition}
\label{sec:image}

Our goal is to collect as much information as possible about potential track
candidates, based on the location and shape of the measured hits in an event.
To accomplish this, the position of each hit is transformed to a
four-dimensional $(\q_R, \sinh\eta, \phi_0, z_0)$ accumulator space of track
parameters.
The accumulator space is not continuous but binned. (Bins are consecutive,
adjacent, non-overlapping equal size intervals of a variable.) 
Ranges and the optimised number of bins corresponding to
track parameters are shown in Table~\ref{tab:bins}.
The transformation is a variant of the well-known Hough
transform~\cite{hough1962method}. In the absence of physical effects
(Sec.~\ref{sec:physics}) the image of a point-like $(r\phi, z)$ hit would be a
well-defined two-dimensional manifold in that space, while the image of a
section-shaped strip hit would be a three-dimensional manifold.

\subsection{Preparation of templates}

The detector models studied here have translational symmetry in longitudinal
($z$) and rotational symmetry in azimuthal ($r\phi$) direction. The
$\phi-\phi_0$ angular difference primarily depends on $\q_R$, while the $z-z_0$
longitudinal difference is mostly a function of $\sinh\eta$
(Fig.~\ref{fig:calib_coll}). These difference distributions further depend on
the shape of the hit cluster (Sec.~\ref{sec:clusters}). For particles with a given $(\q_R, \sinh\eta)$ and
mass, the $(\phi-\phi_0, z-z_0)$ values on a given detector layer populate a
small rectangular area. The dimensions of that area result from the binning of
the track parameters.

With help of numerous simulated particles we determine the populated
area with help of local linear approximations
\begin{gather*}
 \begin{pmatrix} \phi-\phi_0 \\ z-z_0 \end{pmatrix} =
 \begin{pmatrix} \phi-\phi_0 \\ z-z_0 \end{pmatrix}_c +
 \frac{\partial(\phi-\phi_0, z-z_0)}{\partial(\q_R, \sinh\eta)}
 \begin{pmatrix} \Delta \q_R \\ \Delta \sinh\eta \end{pmatrix}
\end{gather*}

\noindent
around the center of each $(\q_R, \sinh\eta)$ bin. In practice the
$(\cdot)_c$ central values and the $\partial(\cdot)/\partial(\cdot)$ Jacobian
is deduced for each bin. The set of these values will be referred to as
templates in the following.

In order to have uniform coverage in all bins, the distribution of simulated
particles is chosen to be constant in $\q_R$, $\sinh\eta$, and $\phi_0$. To
limit fluctuations, normally distributed random variables, used in the simulation
of physics processes (Sec.~\ref{sec:physics}), are limited to values within 3.5
standard deviations (only about 0.05\% lies outside this range).  Altogether $2
\times 10^6$ pions are generated.

\begin{figure}[!b]

 \begin{center}
  \resizebox{\linewidth}{!}{
   \input{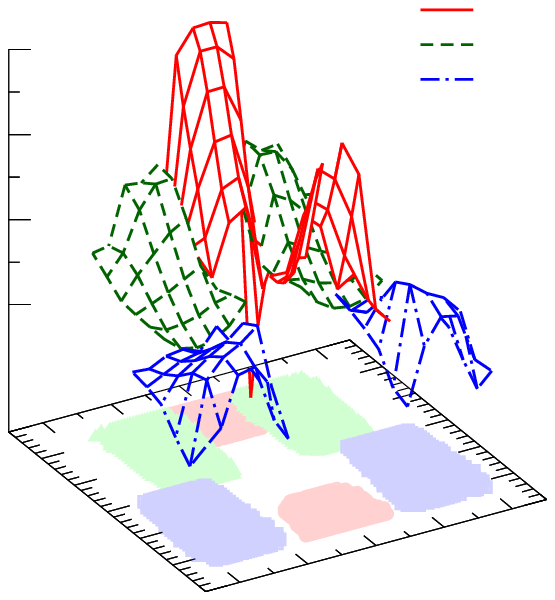}
  }
 \end{center}

 \caption{The role and use of cluster shape information is shown through the
distribution of template values (their width in $\phi$ direction) as a function
of $(\q_R,\sinh\eta)$. The clusters with given $(w_{r\phi},w_z)^q$ shape values
are taken from a given pixel layer. For better discrimination between the three
cluster shapes, the projections of their set of points are also shown on the
base with light colour.}

 \label{fig:shape}

\end{figure}

The role and use of cluster shape information is shown through the
distribution of template values (their width in $\phi$ direction) as a function
of $(\q_R,\sinh\eta)$ in Fig.~\ref{fig:shape}.

The prepared templates are used in two ways. During the early stage of image
transformation they provide a $(\phi_0,z_0)$ accumulator area to increment for
each (pixel) hit, in the case of a given $(\q_R, \sinh\eta)$ bin. Later they
are used to specify a search rectangle on the $(r\phi,z)$ plane of a (strip)
layer for a given $(\q_R, \sinh\eta, \phi_0, z_0)$ bin.

Although detector models with only barrel silicon detectors are studied here,
the above considerations can be adopted to other geometries, such as disks
perpendicular to the beam axis. In that case, the translational symmetry would
be lost and the templates would become more complex by introducing another
dimension, namely the relative $z$ position of the primary interaction with
respect to the longitudinal coordinate of the disk.

\subsection{Image transformation}

The transformation of spatial information to track space proceeds as described
in the following.
First hits on the three innermost layers are dealt with, containing exclusively
pixel hits. For each hit all potential $(\q_R, \sinh\eta)$ accumulator bins
are examined and the corresponding possible $(\phi_0,z_0)$ values, enveloped by
rectangles, are determined. Bins within such $(\q_R, \sinh\eta, \phi_0, z_0)$
area are incremented.
Since we look for tracks with hits on all three innermost layers, only those
accumulator bins are kept which gathered votes from all three layers.

The subsequent layers usually contain strip hits. The method used for the
innermost layers would not be efficient here because there are far too many
accumulator bins to handle.
To this end, the kept accumulator bins are examined, corresponding to
proto-tracks with three counts, obtained in the previous step. Using the bin
coordinates $(\q_R, \sinh\eta, \phi_0, z_0)$ we look for compatible hits by
determining a search rectangle on the $(r\phi,z)$ plane for each layer. For
quick access, and in order to facilitate hit selection, strip hits are in
advance partitioned on an equidistant grid using their $(r\phi,z)$ coordinates.

\begin{figure}[!t]

 \begin{center}
  \resizebox{\linewidth}{!}{
   \input{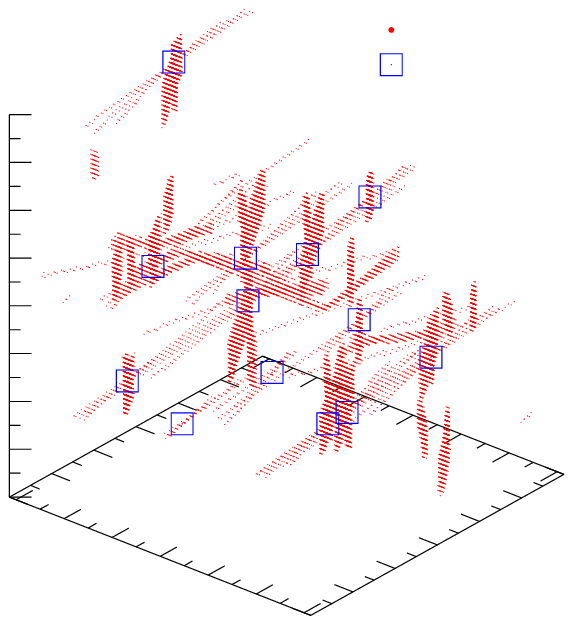}
  }
 \end{center}

 \caption{Distribution of hit images (red dots) in $(\q_R, \sinh\eta, \phi_0)$
accumulator space in the case of a single pp collision for a given $z_0$ value.
The found track candidates are marked with blue boxes.}

 \label{fig:trans}

\end{figure}

The search for compatible hits proceeds outwards. It is advantageous since the
process can be abandoned if some layers provided no compatible hits while they
are still reachable according to the curvature range of the examined bin. In
other words, the number of layers with compatible hits should not be very
different from the number of reachable layers. (We can allow for a few layers
without hits.)

\subsection{Trajectory building}

Track candidates are built using hit images collected in given accumulator bins
(Fig.~\ref{fig:trans}).
Trajectory propagation and track fitting is performed by the extended classical
Kalman filter~\cite{Fruhwirth:1987fm} including prediction, filtering, and
smoothing, with pion mass assumption.
The initial state vector is estimated by placing a helix to the innermost two
hits and using the beamline as a constraint by adding a zeroth point with value
$r\phi=0$, and with an uncertainty of $\sigma_{r\phi} \approx 50\,\mum$. In the
case of an off-centered beam, $\sigma_{r\phi}$ can be increased to properly
contain the interaction region in the transverse plane.

Trajectory building normally proceeds from inside out and considers all hit
combinations recursively by forming branches. At each layer there are usually
multiple hits to add to the existing trajectory. The number of hits
to be considered is especially large for the inner strip layers that would
exponentially increase the number of trajectory branches.

\begin{figure}[!b]

 \begin{center}
  \resizebox{\linewidth}{!}{
   \input{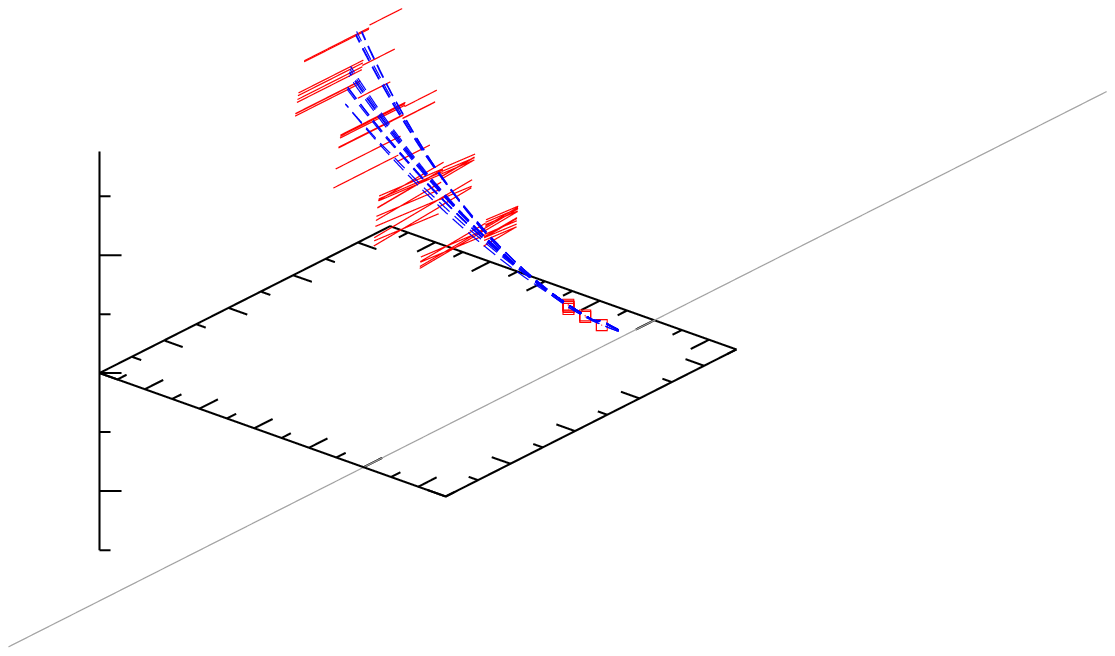}
  }
 \end{center}

 \vspace{-0.5in}
 \caption{Hits (red boxes and line sections) belonging to a given bin in the
accumulator space and trajectories propagated to the outermost detector layer
(blue dashed curves). The beamline is indicated with the gray straight line.}

 \label{fig:hit_families}

\end{figure}

A powerful solution for this problem is the effective detection of unsuitable
(outlier) hits in the busy detector layers. For this purpose, trajectory
building starts with the estimate of the initial state vector at the zeroth
point, which is then propagated through the first three layers. At that point
we already have a good knowledge about the parameters of the track that is
being reconstructed. Instead of going to the next (strip) layer, the trajectory
is at once propagated to one of the potential hits in the outermost detector
layer (Fig.~\ref{fig:hit_families}). During propagation the physical effects of
the crossed layers are duly taken into account but information about their hits
is not used. Next, the outliers in the intermittent omitted layers are detected
in the smoothing step of the deficient trajectories using the smoothed
residual~\cite{Fruhwirth:1987fm}. Hits in the upper 0.5\% tail of the
corresponding $\chi^2$ distribution are discarded.

\begin{figure*}[!t]

 \begin{center}
  \includegraphics[width=\linewidth]{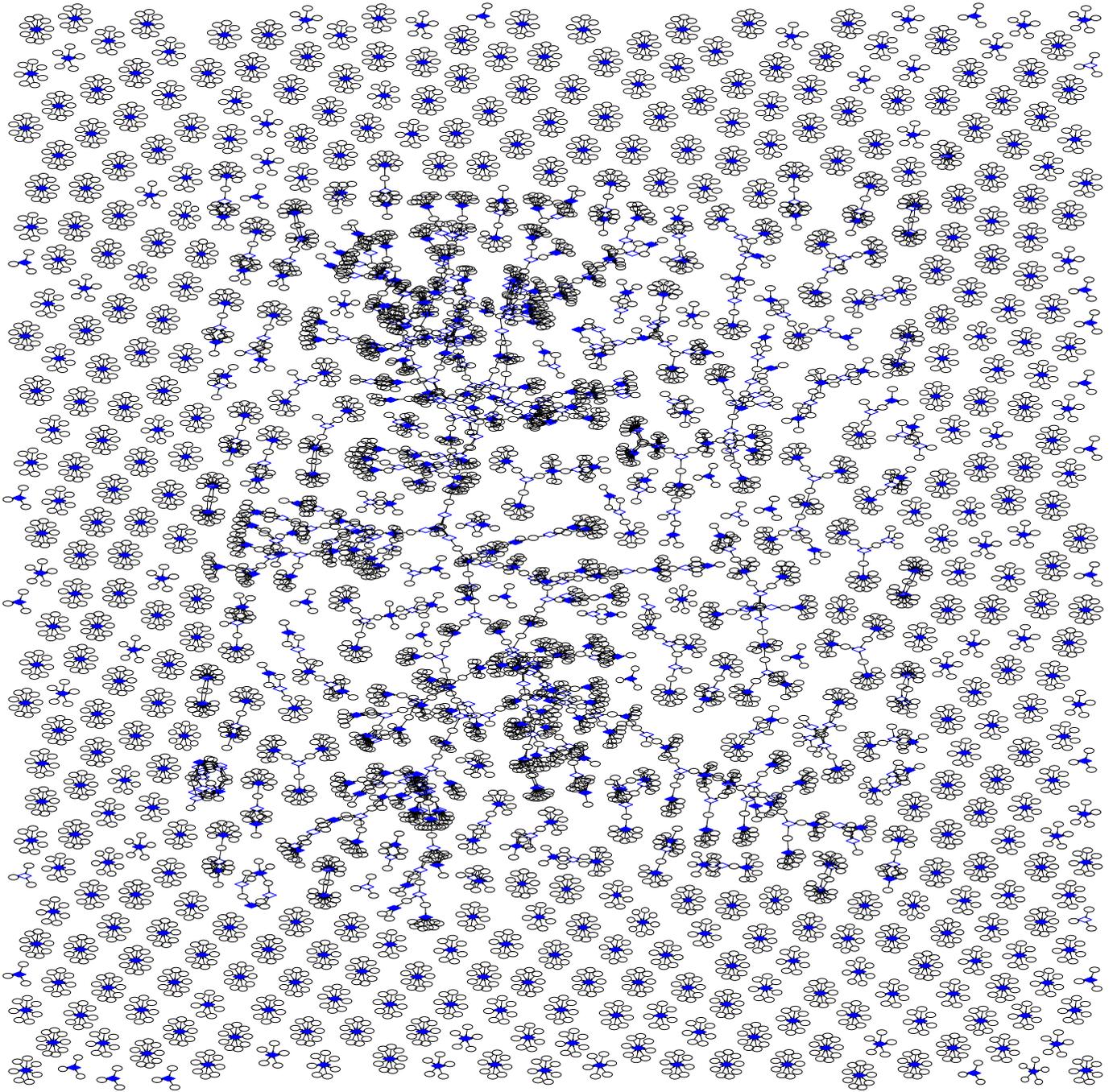}
 \end{center}

 \caption{The bipartite graph $G$ of hits (black ellipses) and track candidates
(blue diamonds) for an event with multiple (40) pp collisions. Directed arrows,
graph edges, show potential hit-to-track candidate assignments.
Filled diamonds indicate true tracks, while open ones show candidates where one
or more hits are not in place.}

 \label{fig:graph_all}

\end{figure*}

Once the list of compatible strip hits is narrowed down, full trajectory
building with the selected hits is performed again. During trajectory building
we can allow for a few (one or two) missing hits. It may mean no hit at all or
too large $\chi^2$ for a given number of degrees of freedom (ndf). If there are
too many missing hits, the process is abandoned. In order not to lose a
noticeable amount of track candidates, but also to have a good selection power,
trajectories in the upper 0.5\% tail of the corresponding $\chi^2$ distribution
are discarded and not developed further (roughly those with $\chi^2 \lesssim
1.7~\mathrm{ndf} + 8.0$ are kept).

\section{Optimal distribution of hits among tracks}
\label{sec:graph}

\begin{figure}[!t]

 \begin{center}
  \includegraphics[width=\linewidth, trim={1.1in 1.07in 1.1in 1.07in},
clip=true]{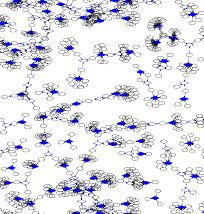}
 \end{center}

 \caption{A small fraction of the bipartite graph $G$ of hits (black ellipses)
and track candidates (blue diamonds) for an event with multiple (40) pp
collisions.  Directed arrows, graph edges, show potential hit-to-track
candidate assignments.
Filled diamonds indicate true tracks, while open ones show candidates where one
or more hits are not in place.}

 \label{fig:graph}

\end{figure}

In the end we have a set of track candidates with the somewhat unusual property
that temporarily several track candidates may share some hits. Our goal is to
resolve these ambiguities, hit confusion, by optimally allocating the hits
among tracks, since all hits must be assigned to not more than one track.
(The tasks is called optimal packing in mathematics.)
The hits and track candidates, and their relations, are best represented by a
bipartite graph $G$. The nodes of $G$ are from two disjoint sets: hits and
track candidates, such that each edge connects a hit to a track candidate
(Figs.~\ref{fig:graph_all} and \ref{fig:graph}). Our goal is to assign all hits
to at most one track, while keeping an eye on the total goodness-of-fit of all
tracks ($\sum\chi^2$) in the event.

First those track candidates are privileged which are very likely real. To
this end, we extract a subgraph by selecting track candidates which have at
least three hits not requested by other candidates, that is, at least three
leaf hit nodes. (A leaf hit node is connected to exactly one track candidate
node.) This subgraph is disconnected (Sec.~\ref{sec:disconnect}), the arising
minigraphs are solved (Sec.~\ref{sec:solve}) individually. Selected tracks and
stored, edges and nodes of the subgraph are removed from $G$.
Next we extract the subgraph containing track candidates with the highest
maximum number of possible hits (usually $n=9$), and their corresponding hits.
Thanks to the high number of hits required, these track candidates are likely
real. Selected tracks are again stored, nodes and edges removed from $G$ as
above. Then the process is restarted with the subgraph of track candidates with
$n-1$ hits, iterating down to the subgraph of track candidates with three hits.

\subsection{Disconnecting a subgraph}
\label{sec:disconnect}

The graphs encountered are usually highly connected. If we would try to
allocate the hits to tracks one by one, the number of trials needed would
explode exponentially with increasing number of nodes. In order to reduce the
complexity of the problem, the graphs should be partitioned into several small
pieces. This can be accomplished by finding some vulnerable components in the
graph whose removal disconnects the graph. Such weak elements are special edges
(bridges) and special nodes (articulation points) whose deletion increases the
number of connected components of the graph. Of course this way some tracks
would lose a hit, but that is only a small price to pay.

\begin{figure}[!b]

 \begin{center}
  \includegraphics[width=\linewidth]{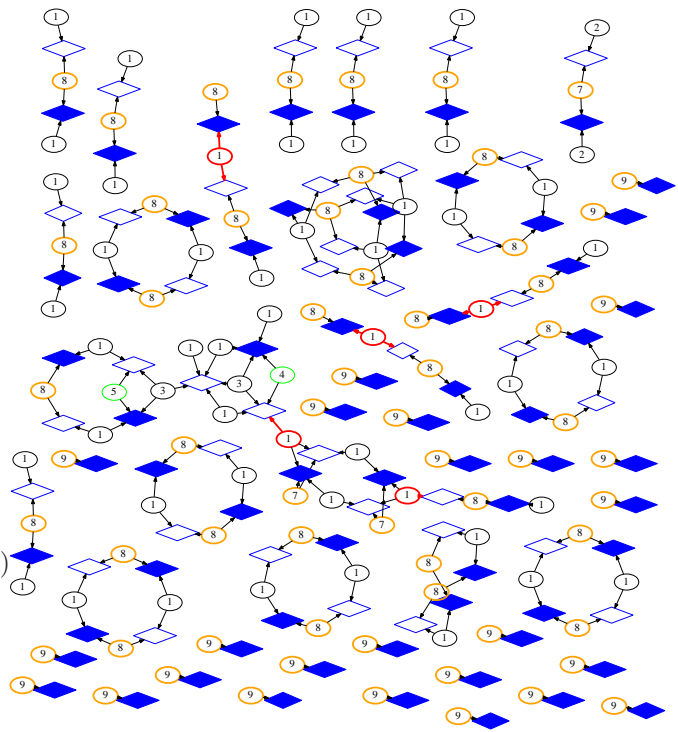}
 \end{center}

 \caption{Example minigraphs obtained after the removing most of bridges and
articulation points from the bipartite graph $G$ of hits and track candidates,
in the case of an event with multiple (40) pp collisions. The thickness and
colour of contracted hits (ellipses) refer the number of hits with identical
role they represent (orange -- 6 or more, green -- 4 or 5, black -- 3 or less),
that number is printed within the corresponding ellipses.
Filled diamonds indicate true tracks, while open ones show candidates where one
or more hits are not in place.
The remaining bridges and articulation points are drawn with thick red arrows
and ellipses, respectively.}

 \label{fig:minigraphs}

\end{figure}

Bridges and articulation points can be found in linear time with help of graph
traversal techniques. The depth-first search is an algorithm for traversing and
searching a graph. One starts at some arbitrary node as the root and explores
as far as possible along each branch before backtracking. The nodes of bridges
and the articulation points are found by requiring that their children nodes do
not have a backedge.
 
In a ``disconnecting'' step the found bridges and articulation points are
removed. During the process new vulnerable elements may come to light, hence
the disconnecting steps are repeated until no new such elements are found.
As a next step, the resulted graph is further partitioned into disjoint graphs.
This task is best accomplished by the flood fill method embedded into the above
detailed traversal technique. The output of the disconnecting step is a large
set of disjoint minigraphs.

In high-pileup pp events there are usually several thousand track
candidates. Their corresponding bipartite graph $G$ and its subgraphs contain
several hundred bridges and up to 50 articulation points. Once the subgraphs
are disconnected, we get couple of thousand minigraphs.

\subsection{Solving a minigraph}
\label{sec:solve}

A minigraph usually has several hits with identical role: they are connected to
the same set of track candidates. (Their number is between 2 and 7 hits for the
detector models studied here.) In the interest of reducing complexity, hits
with identical role are treated jointly, the set of such hits gets
``contracted''.

The number of remaining nodes is usually small (Fig.~\ref{fig:minigraphs}), the
contracted hits can be distributed among tracks by building and solving a
decision tree. The process is similar to exploring decision trees of
deterministic strategy board games, such as chess and go, including their
horizon problem (limited search depth).  What is different here is that our
process is a single-player one.
The optimal hit-to-track assignments are chosen in the following way,
recursively:

\begin{figure}[!b]

 \begin{center}
  \resizebox{\linewidth}{!}{
   \input{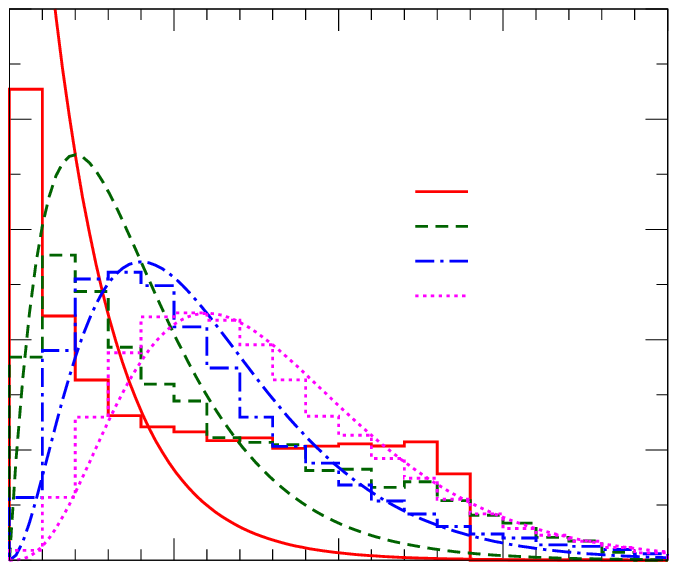}
  }
  \resizebox{\linewidth}{!}{
   \input{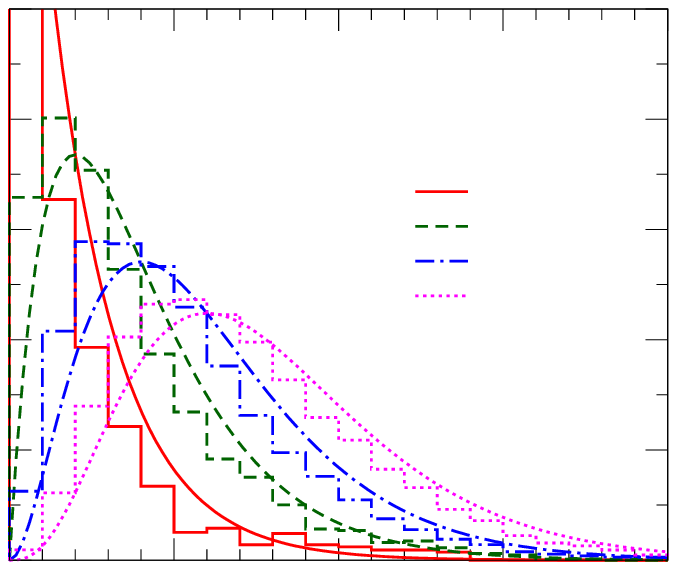}
  }
 \end{center}

 \caption{Comparisons of track-fit $\chi^2$ distributions of track candidates
(top) and of final tracks (bottom) for events with multiple pp collisions.
Distributions from data (histograms) and compared to theoretical expectations
(curves) for tracks with given number of degrees of freedom (ndf).}

 \label{fig:chi2}

\end{figure}

\begin{enumerate}

 \item First the most important, highest ranked available (contracted) hit is
located. Rank is calculated as the product of the number of similar hits (the
number of hits the contracted hit represents) and the number of edges the hit
node has (the number of associated track candidates).
Such a definition gives preference to contracted hits which represent many
particle hits and have a central role in the graph.

 \item The highest ranked hit can be attached to several track candidates, and
these choices are evaluated sequentially and recursively as branches of a
decision tree. After a hit-track assignment is chosen, the track and its hits
are selected, and their nodes and all corresponding edges are removed from the
minigraph.

 \item Track candidate nodes and corresponding edges with too few remaining
hits (less than three) and those with too many missing hits are also removed.

 \item As long as there are nodes left in the minigraph we go back to step 1,
otherwise the actual path of the decision tree is evaluated based primarily on
the amount of hits on selected tracks. If there are two decision trees with
the same amount of hits, the one with lower $\sum\chi^2$ of the selected tracks is
chosen.

\end{enumerate}

In order to save time during solving the decision tree, the selected tracks
candidates are not re-fitted but only the adjusted $\chi^2$ values of their
remaining hits, calculated based on a Kalman-fit using all the initial hits,
are summed and the corresponding ndf values are recalculated.

In the end the decision path with the best score is taken, the selected tracks
and their hits are stored.

\section{Results}
\label{sec:results}

In the computer simulation the interaction region is centered at $(x,y,z) =
(0,0,0)$. In $z$ (beam) direction it is described by a Gaussian distribution
with a standard deviation of $\sigma_z =$ 5~cm. The silicon tracker covers the
pseudorapidity range of $|\eta| < 1.5$.

Inelastic pp collisions at $\sqrt{s} =$ 14~TeV are obtained from the {\sc
Pythia}8~\cite{Sjostrand:2007gs} Monte Carlo event generator (version 219). It
is known to well reproduce the measured momentum spectrum of charged particles
in pp collisions at $\sqrt{s} =$
13~TeV~\cite{Khachatryan:2015jna,Adam:2015pza,Aad:2016mok} with an average
pseudorapidity density of $dN/d\eta \approx 5.5$ near $\eta \approx 0$, as well
as the composition of the most abundant charged particles (pions, kaons,
protons).
Semi-central PbPb collisions at $\sqrt{s_\text{NN}} =$ 5.5~TeV with $dN/d\eta
\approx 1000$ are obtained from the {\sc Hydjet}~\cite{Lokhtin:2005px} Monte
Carlo event generator (version 1.9). It was tuned to match the measured
momentum spectrum of charged particles in the highest energy heavy ion
collisions, as seen for $\sqrt{s_\text{NN}} =$ 5.02~TeV energy central PbPb
collisions~\cite{Adam:2015ptt}.

Physical effects (multiple scattering and energy loss) are simulated according
to the simple models detailed in Sec.~\ref{sec:physics} using the description
of detector materials shown in Table~\ref{tab:setups}.

\begin{figure*}

 \begin{center}
  \resizebox{\textwidth}{!}{
   \input{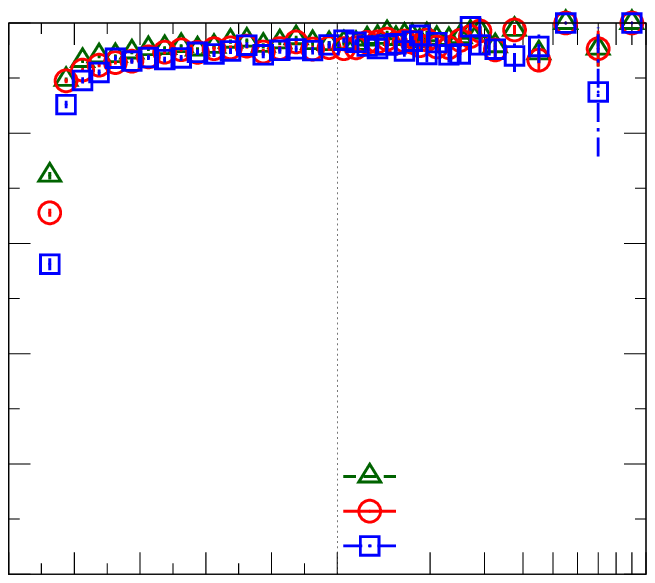}
   \input{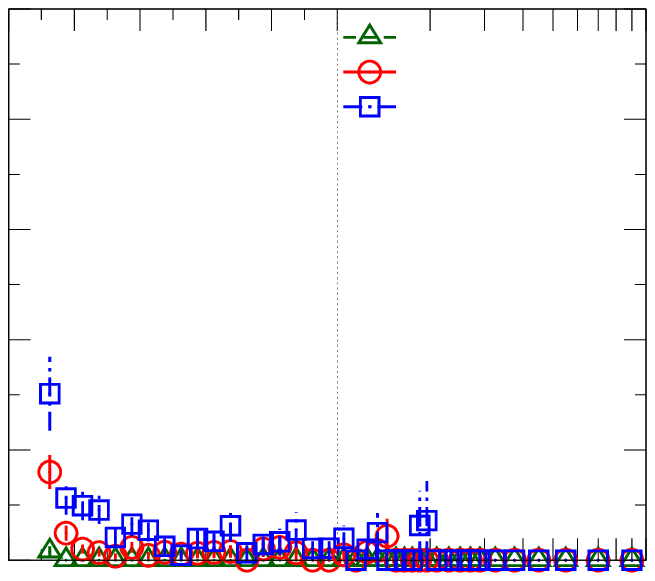}
  }

  \resizebox{\textwidth}{!}{
   \input{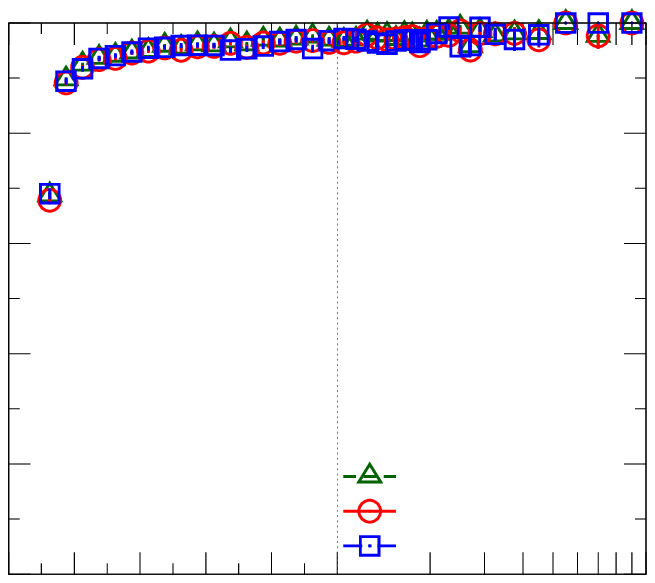}
   \input{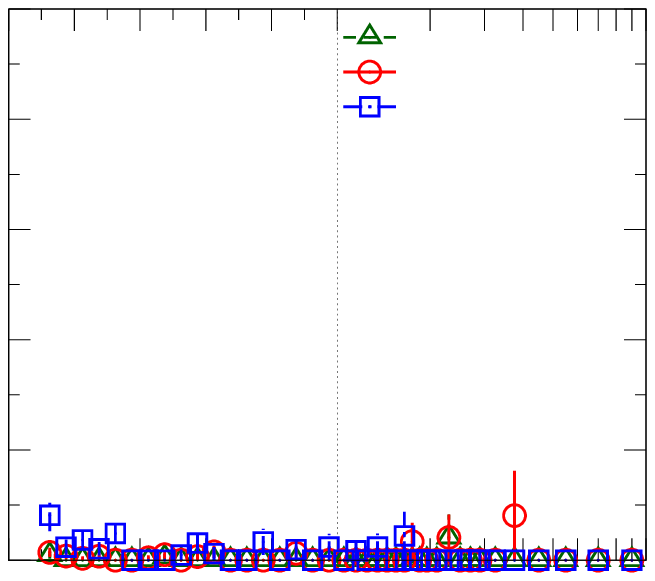}
  }

  \resizebox{\textwidth}{!}{
   \input{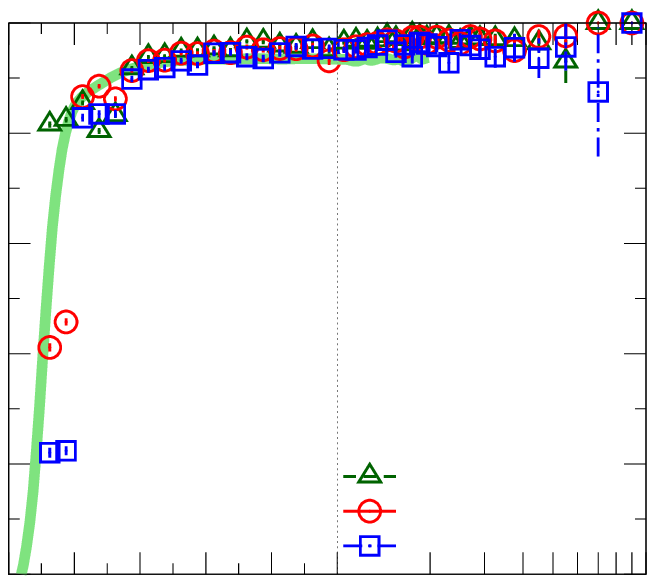}
   \input{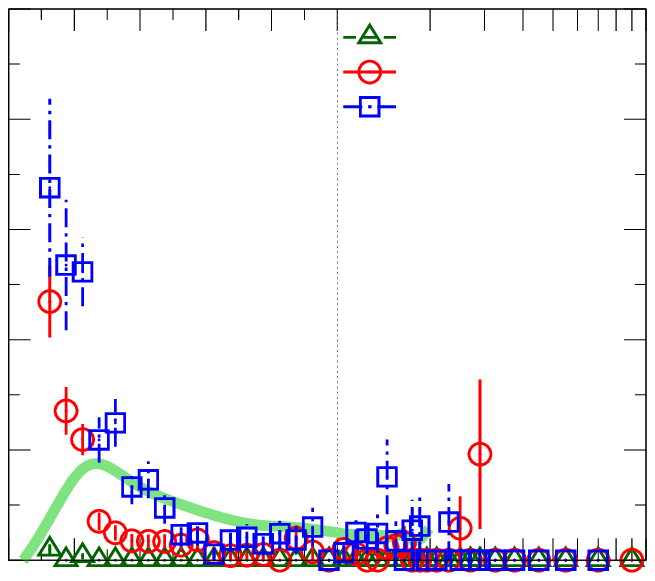}
  }
 \end{center}

 \caption{Performance of the proposed algorithm as a function of transverse
momentum ($\pt)$ of the charged particles. Efficiency (left) and fake track
rate (right) are plotted for Exp A (top), Exp B (middle), and Exp C (bottom).
The values are separately given for single pp (green triangles), multiple (40)
pp (red circles), and semi-central PbPb (blue boxes) collisions. The horizontal
scale is linear in the region 0--1~\GeVc, while it is logarithmic for
1--10~\GeVc. In the case of Exp C, values for single pp events
from~\cite{Sirunyan:2017zmn} are shown for comparison with light green bands.}

 \label{fig:perf_vs_pt}

\end{figure*}

Comparisons of track-fit $\chi^2$ distributions of initial track candidates and
of final tracks for events with multiple pp collisions are shown in
Fig.~\ref{fig:chi2}. Distributions from data and compared to theoretical
expectations for tracks with given number of degrees of freedom.  While track
candidates have distorted $\chi^2$ distributions, those of final tracks are
much closer to the expected curves.

The proposed algorithm was coded in C++ and run on a 3.1~GHz quad-core
computer. The average CPU time needed was on average 13~sec for events with 40
simultaneous inelastic pp collisions.
 
The performance as a function of transverse momentum $\pt$ of the charged
particles is shown in Fig.~\ref{fig:perf_vs_pt}. (A reconstructed track is
considered matched to a simulated one if all their hits correspond to each
other, or if at most one of them is not in place.) Efficiency and fake track
rate are plotted for the three experimental setups. The values are separately
given for single pp, for multiple simultaneous pp, and for semi-central PbPb
collisions. It is clear that for $\pt >$ 0.2\,\GeVc\ the efficiency is above
90--95\% and fake track rate is well below 1\%, independent of collision system
(pp, PbPb) and pileup (1--40). At very low transverse momentum ($\pt <$
0.2\,\GeVc) efficiency drops and fake track rate increases to some 2--4\%.
Both measures show a clear advantage and convincing performance of the proposed
method over those presently used in the highest energy particle physics
experiments, where performance usually decreases with increasing event
multiplicity.

\begin{figure}[!t]

 \begin{center}
  \resizebox{\linewidth}{!}{
   \input{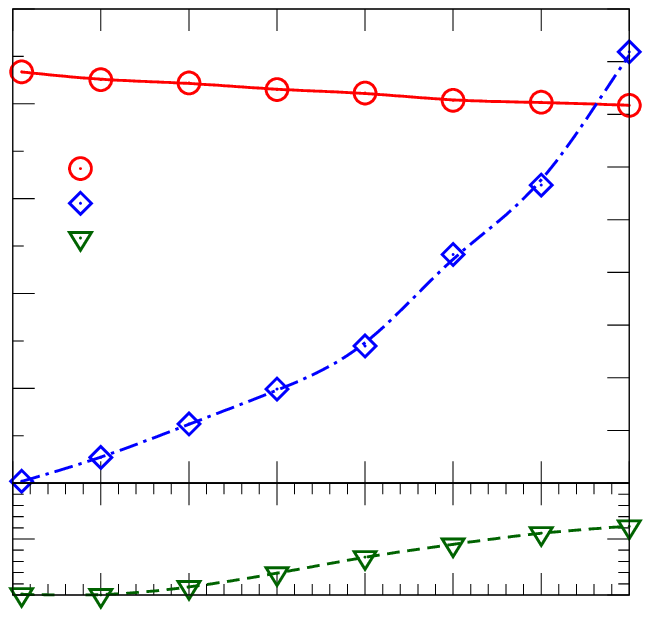}
  }
 \end{center}

 \caption{Performance of the proposed algorithm as a function of the number of
simultaneous inelastic pp collisions. Efficiency (open red circles, top left),
running times (open blue diamonds, top right), and fake track rate (open green
triangles, bottom) values are plotted. Lines are drawn to guide the eye.}

 \label{fig:perf_vs_mu}

\end{figure}

In the case of multiple pp collisions most computing time is spent on the image
transformation, while for PbPb trajectory building takes most of the resources.
Disconnecting and solving the graph is quick in both cases.
Efficiency, running times, and fake track rate values as a function of the
number of simultaneous inelastic pp collisions are shown in
Fig.~\ref{fig:perf_vs_mu}. Efficiency for all charged particles slowly
decreases but stays in the 80--90\% range. The fake track rate starts at the
permille level and stays under a percent.

\section{Summary}
\label{sec:summary}

A combination of established data analysis techniques for charged-particle
reconstruction was presented. The method follows a global approach and uses all
information available in a collision event. It employs image transformation
based on precomputed templates taking advantage of the translational and
rotational symmetries of the detectors. Track candidates and their
corresponding hits form a usually highly connected network, a graph. The graph
is partitioned into very many minigraphs by removing a few of its vulnerable
components, edges and nodes. The hits of the subgraphs are distributed among
the track candidates by solving a deterministic decision tree.

Tests using simplified computer models of LHC silicon trackers show that
efficiency and purity of track reconstruction are excellent and the timing of
the proposed method is reasonable, both in simultaneous proton-proton
collisions (high pileup), and in single heavy-ion collisions at the highest
available energies.

\begin{acknowledgement}

The author wishes to thank to S\'andor Hegyi and Andr\'as L\'aszl\'o for
helpful discussions.
This work was supported by the Swiss National Science Foundation (SCOPES
152601), and the National Research, Development and Innovation Office of
Hungary (K 109703).

\end{acknowledgement}

\bibliographystyle{epj}
\bibliography{denseTracking}

\end{document}